\documentstyle[12pt]{article}

\newcommand{\lesssim}{
 \mathrel{\setbox0=\hbox{$<$}\raise0.6ex\copy0\kern-\wd0
 \lower0.65ex\hbox{$\sim$}}}
\newcommand{\gtrsim}{
 \mathrel{\setbox0=\hbox{$>$}\raise0.6ex\copy0\kern-\wd0
 \lower0.65ex\hbox{$\sim$}}}

\begin{document}

\title{Soft-hard interplay and factorization for baryon production
in the target fragmentation region in ep collisions}

\author{ Leonid Frankfurt\thanks{\noindent On leave of absence from the
          St.Petersburg Nuclear Physics Institute, Russia.}\\
         School of Physics and Astronomy\\
         Raymond and Beverly Sackler Faculty of
         Exact Sciences\\
         Tel Aviv University, Tel Aviv, Israel
\\[0.3cm]
         Werner Koepf\\
         Department of Physics, The Ohio State University\\
         Columbus, OH 43210, USA
\\[0.3cm]
         Mark Strikman\thanks{\noindent Also at St.Petersburg Nuclear Physics
          Institute, Russia.}\\
         Department of Physics, Pennsylvania State University\\
         University Park, PA 16802, USA
\\[0.6cm]}

\date{February 2, 1997}

\maketitle

\vspace{-20.0cm}
\begin{flushleft}
\tt DESY-97-027, OSU-97-0202 \\ February 1997
\end{flushleft}

\newpage

\begin{abstract}
We discuss baryon production in the nucleon fragmentation region in deep
inelastic scattering.
 The dependence of
the nucleon spectra on Bjorken-$x$ is evaluated within the parton model and in
QCD, and ways to look for the break-down of the DGLAP approximation via
baryon production are suggested.  We argue
 also that the leading neutron
production in these small $x$ processes is rather insensitive to the
pion parton densities of the nucleon.
\end{abstract}

\section{Introduction}

Current HERA experiments, both in the collider and fixed target modes,
allow to take a fresh look at the target fragmentation region in a wide
range of $x$.

It has been discussed for a long time within the framework of the parton model that
 at large $x$ properties of the target fragmentation region should depend on $x$, see e.g.
Ref. \cite{FS77}.
 On the contrary for small $x$
the Yang limiting fragmentation  is expected i.e.
the baryon differential multiplicities should be
 similar to those in soft hadron-nucleon interactions.
 Really within the parton model hadron radiation is  a random process
where transverse size of initial configuration diffuses to the
transverse size of the target and memory on  the hard physics  in the
hadron final
states is restricted by the photon fragmentation region which form
small part of rapidity space at fixed $Q^2$ when $x\rightarrow 0$.

This diffusion is especially enhanced in PQCD as a result of infrared
slavery, i.e., from the infrared pole in the invariant charge. As a
result of this diffusion, the interaction  with the target
 is always given by non-perturbative QCD, and it is independent of
the projectile. At the same time, it is expected that with
increase of  $x$
 the spectrum of baryons should drop faster at large   $z$, where
$z$ is the Feynman-$x$ of the produced baryon, see Eq. (\ref{zdef}) below.

We reevaluate these expectations by taking into account effects of
the gluon bremsstrahlung and consider also
new opportunities for testing the small
$x$ dynamics of the strong interaction by means of studies of
long range correlations in rapidity space
in deep inelastic scattering.  This is now practical with HERA's
large acceptance detectors.  In particular, we will argue that the
lack of correlations between hadron production in the current
 and target fragmentation regions, which is  characteristic
for the DGLAP evolution equations,
may break down if non-linear effects are really important
in the evolution.  These non-linearities could be
observed simply by switching from
inclusive measurements to semi-inclusive studies involving the production
of hadrons at central rapidities.

\section{Parton model expectations}

First, let us summarize the parton model expectations. It
is convenient to consider the scattering in the Breit frame,
where the colliding proton and the virtual photon have four-momenta
$p_N=(P,P,\vec 0_t)$ and $q_{\mu}=(0,-2xP,\vec 0_t)$, respectively.
After the violent collision, the photon
removes a parton with momentum $xP$ and ``turns it around'', while
the spectator system of partons with momentum $(1-x)P$
fragments into hadrons in the proton fragmentation region.

It is convenient to define the Feynman-$x$ for the produced baryon,
\begin{equation}
z~=~{p_{h} \over (1-x)P} \ ,
\label{zdef}
\end{equation}
as well as the light-cone fraction $\beta$ of the initial proton momentum
carried by this baryon,
 \begin{equation}
\beta~=~{p_{h} \over P}~=~z(1-x) \ .
\label{beta}
\end{equation}

Correlations between hadrons produced in the soft
 interactions are
 short-range
 in rapidity, i.e., $|\Delta y| \le 2$. Therefore, one expects that in the
limit of small $x$, when the distance in rapidity between the removed
parton carrying the light-cone fraction $x$ and the
leading partons, which form the leading hadron
 exceeds several units
 correlations should disappear.

Suppose  the leading hadron carries
 a light-cone fraction of the residual system's momentum of
Feynman-$z$.  Then, the hadron yield,
\begin{equation}
f(z,p_t,x,Q^2)~\equiv~
{1 \over \sigma_{tot}(\gamma^* p)}\,
{d \sigma(\gamma^* +p \rightarrow h +X) \over dz\,dp^2_t}
\ ,
\label{scaling}
\end{equation}
should be independent of $x$, and be
the same as for real photon or hadron projectiles.
This is just an analog of Yang limiting fragmentation observed
in hadron-hadron collisions.  In the limit of $z \rightarrow 1$,
this relation follows from triple Regge analysis
\cite{Mueller,Kancheli}.  Here, the same triple Reggeon diagrams enter
in the inclusive cross section for different projectiles.
The only difference arises from the coupling
of the Pomeron to the projectile, which cancels out in the ratio to
the total cross section for $\gamma^*(h)p$ scattering, and hence it is
canceled also in the
ratio in Eq. (\ref{scaling})\footnote{Discussion of the diffractive case, i.e,
$h=p$ and $z\rightarrow 1$, requires special treatment, see review in Ref.
\cite{AFS}.}.

The violation of limiting fragmentation may occur due to screening effects
(multi-Pomeron exchanges), which are naturally much smaller in the case of $\gamma$ or
$\gamma^*$ projectiles than for nucleon projectiles. These effects
would lead to a certain enhancement of the spectra of the leading
particles for the case of projectiles interacting with a smaller
cross section, and to a violation of limiting fragmentation
at super-high energies due to an increase
of the effective cross section of projectile-nucleon interaction.
 Indication of such a break-down of factorization for
proton production in the triple Pomeron limit was observed at the Tevatron
collider, see e.g. Ref. \cite{Dino}.

With increasing $x$, the factorization relation in Eq. (\ref{scaling})
is expected to break down.
First, the triple Pomeron piece disappears, i.e.,
the ladder becomes too short in rapidity to build the ${1 \over 1-z}$
behavior. Next, at $x \ge 0.2$ the main contribution to DIS starts
to originate from the scattering off the target's valence quarks.
But the QCD counting rules indicate \cite{FS77} that, in this limit,
\begin{equation}
f(z,p_t,x,Q^2)~\propto~(1-z) \ .
\end{equation}
At sufficiently large $x$ (probably $x\ge 0.5$), DIS selects
scattering from the
minimal Fock space configuration $|3 q\rangle$. The
analysis of the leading perturbative QCD diagrams,
which give the dominant contribution in the $x \rightarrow 1$
limit, indicates that the main contribution stems from configurations
where two spectator quarks carry large relative momenta, i.e.,
${x_1 \over x_2} \gg 1$  or ${x_1 \over x_2} \ll 1$, where $x_1$ and
$x_2$ are the light-cone fractions of the spectator quarks
($x_1+x_2=1-x$) \cite{FS81}. As a result, the independent
fragmentation of two spectator quarks becomes more and more important.
Also, for joint fragmentation of both quarks, the
relative chance of producing an excited baryon state increases.
Hence, in the limit of large $x$, we expect a graduate decrease of
$f(z,p_t,x)$ at large $z$
and an increase of the yield of excited baryon states \cite{FS77,FS81}.

\section{QCD radiation effects}

For inclusive processes, the major modification to the target
fragmentation is due to QCD radiation in the initial state. Really,
according to the QCD evolution
at large $Q^2$, if the $\gamma^*$ interacts with a parton at a given $x$,
latter parton originates from a parent parton  at a  softer
resolution scale $Q^2_0$ with $\tilde x > x$.
As soon as we focus on the production of hadrons at large $z$ and small
enough $k_t$ (where $k_t^2 \ll Q^2_0$), we can neglect the fusion of
gluons (quarks) emitted in the evolution from ($\tilde x, Q^2_0$)
to ($x, Q^2$).  The reason is that these partons
have transverse momenta larger than $\sqrt{Q^2_0}$.
Hence, the simple evolution equations are valid, i.e.,
\begin{equation}
\phi^i(x,Q^2, \beta)\,q_i(x,Q^2)~=~
\int_{ x}^1 d\tilde x\int_{Q_0^2}^{Q^2} d\ln k_t^2~
V^{i,j}\!\left({\tilde x \over x}, {Q^2 \over k_t^2} \right) \,
\phi^j(\tilde x,Q_0^2,\beta)\,q_j(\tilde x,k_t^2)
\ .
\label{evol}
\end{equation}
Here $i$ and $j$ label the parton flavors, $\phi_i(x,Q^2,\beta)$
is the fragmentation function for a
residual system with parton $i$ removed, and $V^{i,j}$
is the hard blob leading to the standard kernel of the
DGLAP evolution equation for inclusive scattering.
The quantity $\beta={z\over (1-x)}$, which was
defined in Eq. (\ref{beta}), is the light-cone fraction of the momentum of the
{\it initial } nucleon carried by the spectator baryon. Conservation of
$\beta$ in the evolution equation reflects the spectator origin of the
leading baryon.
 Within the discussed above approximations
 $Q^2$ dependence of $\phi^i(x,Q^2)$
in Eq.(\ref{evol}) is due to the change of $x$ of the "parent" quark
at $Q_0^2$ scale.

As long as $\tilde x $ satisfies the condition
\begin{equation}
\tilde x~<~x_{diff}~\sim~10^{-2} \ ,
\end{equation}
the factorization expectations of Eq. (\ref{scaling}) should hold.
However, for fixed $x$, with increase of $Q^2$ the essential $\tilde x$
increase, and at some $Q^2$ this inequality should be violated.
Hence, in PQCD, we expect that for fixed $x$ with increase of $Q^2$
the factorization relation would gradually
break down leading to a decrease of $f(z,p_t,x,Q^2)$ at large $z$.

Very recently, ZEUS has released \cite{ZEUS} first data
on the production of neutrons in the reaction
\begin{equation}
e+p~\rightarrow~e+n+X
\label{en}
\end{equation}
in the target fragmentation region for
\begin{equation}
\langle x \rangle \approx 10^{-3}~~\mbox{and}~~Q^2 \ge 10~\mbox{GeV}^2 \ .
\label{kinem}
\end{equation}
The authors did not explicitly check factorization by comparing their
data with real photon data on neutron production or with data on
scattering of circulating protons off the gas in the vacuum tube.
However, qualitatively,
the data seem to be consistent with factorization and a weak ($x,Q^2$)
dependence of the neutron multiplicity. Such comparison
high precision tests of factorization since many detector parameter
uncertainties would cancel out in the relative measurements.

\section{Break-down of factorization in semi-inclusive processes as a test
of the DGLAP approximation at small $x$}

In the kinematics of sufficiently small $x$ in DIS, where the QCD
evolution equation
 should be violated, a contribution of multi-Pomeron exchanges
may appear significant if the rapid increase of the parton
distribution with decreasing $x$ would not be stopped through
the diffusion of small configurations to the soft scale.
Indeed, if we write $\sigma_{tot}$ as a sum of
diffractive (rapidity gap) events and the inelastic contribution,
\begin{equation}
\sigma_{tot}~=~\sigma_{inel}+\sigma_{diff} \ ,
\end{equation}
and assume that only double Pomeron exchanges are important, we
obtain, using AGK cutting rules \cite{AGK}, that single multiplicity
and double multiplicity events have the cross sections
\begin{eqnarray}
\sigma_{single-mult}&=&\sigma_{tot}-3\,\sigma_{diff} \ ,
\\
\sigma_{double-mult}&=&2\,\sigma_{diff} \ ,
\end{eqnarray}
respectively.  Hence,
\begin{equation}
{\sigma_{double-mult}\over \sigma_{single-mult}}~=~
{2\,\sigma_{diff}\over {\sigma_{tot}- 3\,\sigma_{diff}}}~=~0.55-1.0 \ ,
\end{equation}
where we take the diffraction fraction to be 15--20\%.  For
diagrams where two Pomerons are cut,
the spectrum of leading nucleons (with $z \ge 0.5$) should be
reduced substantially similar to the reduction of the spectrum of
leading nucleons in the $p+A \rightarrow N+X$ reaction when the
incoming nucleon interacts with two nucleons in the target.

Substantial fluctuations in the differential multiplicities
of particles produced in the central rapidity range were observed
at HERA for the soft
 $\gamma p$ scattering \cite{H1}. AGK cutting rules
have predicted such fluctuations. Models which include
multi-Pomeron exchanges can quantitatively describe the observed
multiplicity fluctuations, for a recent summary see Ref. \cite{Bopp}.
Thus, it is quite likely that multi-Pomeron exchanges contribute
significantly to the large multiplicity tail of these distributions.

Hence, we suggest to measure the multiplicity of the leading neutrons
as a function of the multiplicity of the hadrons produced at central
rapidities. This   quantity can be defined as:
\begin{equation}
g(z,k_t,x, Q^2, N_h(y_1, y_2))~\equiv~{1 \over
 N(x, Q^2, N_h(y_1, y_2))} {d N (x, Q^2, z, k_t, N_h( y_1, y_2)) \over
{d z \over z} d^2k_t} \ ,
\end{equation}
where $N_h(y_1, y_2)$ is the multiplicity of hadrons
produced in the rapidity interval $y_1 \le y \le y_2$, and
$N(x, Q^2, N_h(y_1, y_2))$ is the number of DIS events with multiplicity
$N_h(y_1,y_2)$.
Screening effects would lead to a decrease of $g(z,k_t, N_h(y_1,y_2)$
for large $z$ if one selects events with
particle multiplicity in the central rapidity range satisfying the inequality
\begin{equation}
N_h(y_1, y_2)~\ge~2\,\langle N_h(y_1, y_2)\rangle \ .
\end{equation}
Obviously, no such correlations are expected in the framework of the DGLAP
evolution equations.

\section{Factorization and the possibility of extracting the pion structure
function from the $e+p\rightarrow e+n+X$ reaction}

It was suggested in a number of papers, see e.g. Refs.
\cite{Julich,Kopeliovich,Levman}, that the HERA collider data on the
process of Eq. (\ref{en}) could be used to measure the pion structure
function at small $x$. Approximate factorization
characteristic of soft QCD processes makes this
extreemly difficult (see however discussion in the end of the paragraph)
since, in this limit, scattering off the Pomeron dominates.
Scattering of any small $x$ partons will lead to essentially the same
spectrum of leading neutrons as does hadron-proton scattering.
Hence, the spectrum of leading neutrons
would be fitted well by a parameterization corresponding to
the sum of $PR_1R_2$ triple Reggeon terms, where $R_i=\pi, \rho,...-$reggeons,
see e.g. Ref. \cite{Field}. Note that in
the high-energy limit, which we discuss, the contribution of $R_1R_2R_3$
terms is negligible\footnote{Note that the $\rho$-Reggeon
has no simple connection to the $\rho$-meson exchange used in models
of the low-energy NN interaction, see e.g. Ref. \cite{Machlid}, since
the effective spin of the $\rho$-Reggeon is $\approx 1/2$ for small $t$
and not 1.
The same is true for pion exchange away from $t \sim 0$. For example,
for $-t \sim 0.2-0.3~\mbox{GeV}^2$, which gives the dominant contribution in
pion models of the nucleon's antiquark sea \cite{FKS}, the pion's effective
spin is $\alpha_{\pi}(t)=(t-m^2_{\pi})+\alpha't\approx -(0.2-0.3)$.}.
At the same time, there is no simple way to distinguish, in the
($x,Q^2$) limit of Eq. (\ref{kinem}) which was studied by ZEUS,
the scattering of the  virtual photon off pions belonging to the nucleon's
$|\pi N\rangle$ and $|\pi \Delta\rangle$ components from scattering
off other components in the nucleon's
wave function.

To estimate the contribution of the scattering off the pion field to the
proton's structure function $F_2(x, Q^2)$, we use our analysis \cite{FKS}
of the nucleon's antiquark distributions at $x \ge 0.15$.  The latter
provides lower limits
on the slopes of the $\pi NN$ and $\pi N\Delta$ form factors.  Note that
the lower limit corresponds to the {\it maximal} possible contribution of the
pion field. It was demonstrated in Ref. \cite{FKS}
that, at small $x$ and especially with increasing $Q^2$, the pion contribution
should be a relatively small fraction of the sea's parton density.
It is straightforward to extend this analysis to the even smaller $x$
of the ZEUS experiment \cite{ZEUS}. We find that, though in this model the
contribution of scattering off the pion field may
exceed $50 \%$ for $Q^2 \le 4~\mbox{GeV}^2$, it cannot
exceed $30-35 \%$ of $F_2(x,Q^2)$ for the kinematics given by
Eq. (\ref{kinem}), see Fig. 1.  There, we show results from fits
to two experiments which measured the pion's parton distributions
via Drell-Yan scattering. The fit to the NA24 data \cite{na24}
leads to a value of the ratio $F_{2\pi}(x,Q^2)/F_2(x,Q^2)$
close to 1 for $x \sim 10^{-3}$ and $Q^2=10~\mbox{GeV}^2$,
while the NA10 fit \cite{na10} corresponds to a ratio of about $0.3$. So,
the results of the calculation based on the NA24 fit
can be considered an upper limit for the pion contribution to
$F_2(x, Q^2)$.

Besides, we want to stress that
since the average relative distances in the $|\pi N(\Delta)\rangle$
configuration are small ($\le 1 fm$) \cite{FKS},
screening effects, which we neglected in our model calculation of
the pion contribution to $F_2$, would further
reduce this contribution.

The main contribution to $F_2(x,Q^2)$ in the small $x$
and  $Q^2 \ge 10~\mbox{GeV}^2$ range comes thus
from the scattering off the gluon field, which,
at low $Q^2$ resolution, originates predominantly from gluon emission
from the valence quarks as well as non-perturbative gluons, and hence
does not belong to the pions. However, based on the factorization argument,
we expect that these configurations give a regular, unsuppressed
contribution to the neutron spectrum.

Within models which explicitly include pion degrees of freedom to the
nucleon's wave function,
scattering off a number of components, $|\pi^+n\rangle$, $|\pi^0 p\rangle$,
$|\pi^-\Delta^{++}\rangle$, $|\pi^0\Delta^{+}\rangle$ and
$|\pi^+\Delta^0\rangle$, is important, see Fig. 2.
Only about $50 \%$ of the final states result
in the production of neutrons. Hence, the total multiplicity of neutrons in
DIS due to the pion mechanism cannot exceed $15 \%$, though for typical
inelastic (non-diffractive) events one expects that the baryon multiplicities
of protons and neutrons are about equal and close to  $50 \%$ (where we
neglected the small correction due to the production of strange baryons).
This expectation is consistent with bubble chamber neutrino
data, which were analyzed, for instance, in Ref. \cite{Lex}.

To really measure the pion structure function,
one would have to extrapolate to the pion pole,
which requires measurements for $-t \sim m_{\pi}^2$. Since
\begin{equation}
-t~=~{m_N^2\,(1-x)\,(1-x-z)^2\over z}~+~{k_t^2\,(1-x)\over z} \ ,
\end{equation}
only the region of $z\ge 0.9$ and $k_t \le m_{\pi}$ could be used
for such measurements.
The current angular and momentum resolution of the ZEUS detector \cite{FNC}
is approaching requirements necessary for such studies.
However counting rates in this region  would be quite low.

Note, also, that even interpolation to the pion pole for small $t$
would not be simple due to screening effects. These effects
lead to a non-zero value for the cross section
at $t=0$, though in a naive pion exchange
model this cross section would be proportional to $t$.
This situation is analogous to the one encountered
in low-energy charge exchange reactions where it causes substantial
problems for an accurate  determination of the $g_{\pi NN}$ coupling
constant, for a recent discussion see Ref. \cite{Ericson}.

To summarize, a study of leading baryon production in deep inelastic
scattering can shed new light on the interplay of soft fragmentation
dynamics and the properties of the nucleon's parton wave function.
It can furthermore
provide an effective probe of the deviations from the DGLAP picture of
small $x$ deep inelastic scattering.

\section*{Acknowledgments}

L.F. and M.S. would like to thank DESY's theory division for its hospitality
during the time when this analysis was carried out. We also like to thank the
members of the ZEUS forward neutron calorimeter group for a number of
useful discussions. This work was supported in part
by the Israel-USA Binational Science Foundation under Grant No.~9200126, by
the U.S. Department of Energy under Contract No. DE-FG02-93ER40771, and by the
National Science Foundation under Grants Nos. PHY-9511923 and PHY-9258270.
\newpage

FIGURE CAPTIONS.

 Fig.1. The relative contribution of scattering off the virtual
pion cloud to the proton's structure function $F_2(x, Q^2)$.  Results are
shown for various $Q^2$ and for the NA24 \protect\cite{na24} and NA10
\cite{na10} pion structure functions.  For further details see
Ref. \cite{FKS}.

Fig.2. The relative contributions of the various virtual
$p\rightarrow B\pi$ processes to $F_2^\pi(x,Q^2)$, the
contribution to the proton's structure function from scattering off its
virtual pion cloud.  Results are shown for $Q^2=10$ GeV$^2$ and for
the NA24 \cite{na24} pion structure functions.

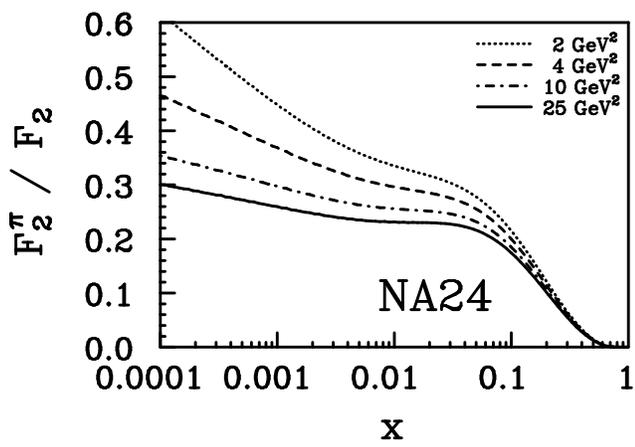
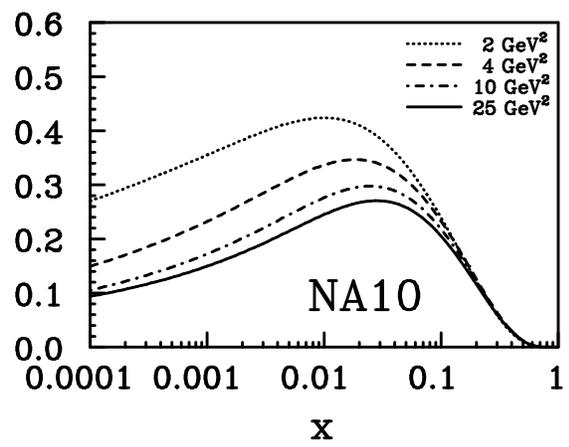
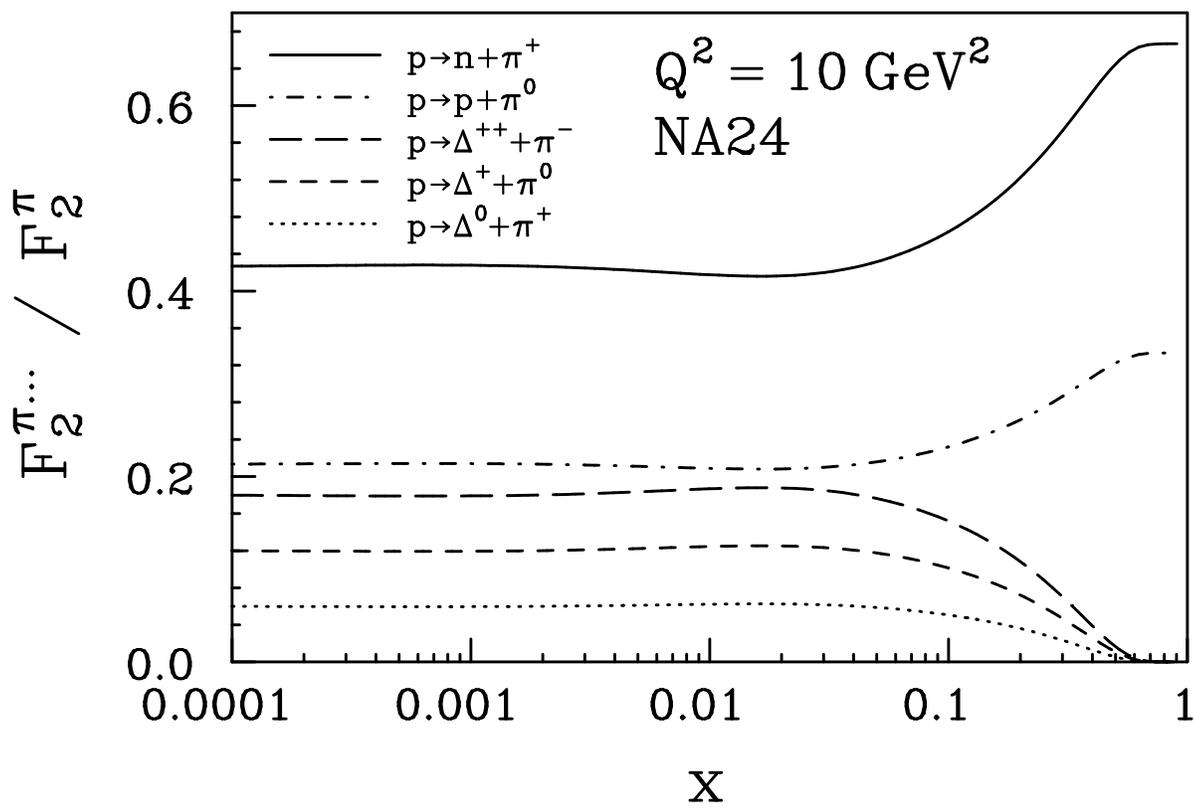
\end{document}